\documentclass{aa}

\usepackage[varg]{txfonts}
\usepackage{siunitx}
\usepackage[version=4]{mhchem}
\usepackage{multirow}
\usepackage{graphicx}
\usepackage[caption=false]{subfig}
\usepackage[strict]{changepage}
\usepackage{xcolor}
\usepackage{ulem}
\usepackage[colorlinks=true,citecolor=blue]{hyperref}
\usepackage{float}
\usepackage{amssymb}
\usepackage{amsmath}
\usepackage{lscape}
\usepackage{mathtools}
\usepackage{natbib}
\usepackage{pdflscape}
\usepackage{gensymb}

\defcitealias{zhou2023}{Paper~I}   

\DeclareSIUnit\jansky{Jy}

\newcolumntype{L}[1]{>{\raggedright\let\newline\\\arraybackslash\hspace{0pt}}m{#1}}
\newcolumntype{C}[1]{>{\centering\let\newline\\\arraybackslash\hspace{0pt}}m{#1}}
\newcolumntype{R}[1]{>{\raggedleft\let\newline\\\arraybackslash\hspace{0pt}}m{#1}}

\newcommand{\mystix} {MYStIX}

\begin{document}

\title{Self-similar cluster structures in massive star-forming regions: Isolated evolution from clumps to embedded clusters}

\author{J. W. Zhou\inst{\ref{inst1}} 
\and Pavel Kroupa\inst{\ref{inst2}}
\fnmsep \inst{\ref{inst3}}
\and Sami Dib \inst{\ref{inst4}}
}
\institute{
Max-Planck-Institut f\"{u}r Radioastronomie, Auf dem H\"{u}gel 69, 53121 Bonn, Germany \label{inst1} \\
\email{jwzhou@mpifr-bonn.mpg.de}
\and
Helmholtz-Institut f{\"u}r Strahlen- und Kernphysik (HISKP), Universität Bonn, Nussallee 14–16, 53115 Bonn, Germany \label{inst2}\\
\email{pkroupa@uni-bonn.de}
\and
Charles University in Prague, Faculty of Mathematics and Physics, Astronomical Institute, V Hole{\v s}ovi{\v c}k{\'a}ch 2, CZ-180 00 Praha 8, Czech Republic \label{inst3}
\and
Max Planck Institute f\"{u}r Astronomie, K\"{o}nigstuhl 17, 69117 Heidelberg, Germany \label{inst4}\\
\email{sami.dib@gmail.com}
}

\date{Accepted XXX. Received YYY; in original form ZZZ}

\abstract
{We used the dendrogram algorithm to decompose the surface density distributions of stars into hierarchical structures. These structures were tied to the multiscale structures of star clusters. 
A similar power-law for the mass-size relation of star clusters measured at different scales suggests a self-similar structure of star clusters.
We used the minimum spanning tree method to measure the separations between clusters and gas clumps in each massive star-forming region. 
The separations between clusters, between clumps, and between clusters and clumps were comparable, 
which indicates that the evolution from clump to embedded cluster proceeds in isolation and locally, and does not affect the surrounding objects significantly. By comparing the mass functions of the ATLASGAL clumps and the identified embedded clusters, we confirm that a constant star formation efficiency of $\approx$ 0.33 can be a typical value for the ATLASGAL clumps.}


\titlerunning{Isolated evolution from clump to embedded cluster}
\authorrunning{J. W. Zhou, Pavel Kroupa, Sami Dib}

\maketitle 

\section{Introduction}


Massive star-forming regions (MSFRs) represent a major mode of star formation within the Galaxy \citep{Lada2003-41,Zinnecker2007,Motte2018,Dib2023-524}. Young stellar clusters within these regions can serve as progenitors for open clusters. It has been suggested that most if not all observed stars originate from embedded clusters \citep{Kroupa1995a-277, Kroupa1995b-277, Lada2003-41,Kroupa2005-576,Megeath2016-151, Dinnbier2022-660}. Measurements of the internal structure and dynamics of young clusters and star-forming regions are necessary to fully understand the process of their formation and dynamical evolution. 


The project called Massive Young Star-Forming Complex Study in Infrared and X-ray \citep[\mystix;][]{Feigelson2013-209} examines 20 nearby (d<3.6~kpc) MSFRs using a combination of archival {\it Chandra} X-ray imaging, 2MASS+UKIDSS near-IR (NIR), and {\it Spitzer} mid-IR (MIR) survey data.
The MYStIX regions include the Orion nebula, the Flame nebula, W~40, RCW~36, NGC~2264, the Rosette nebula, the Lagoon nebula, NGC~2362, DR~21, RCW~38, NGC~6334, NGC~6357, the Eagle nebula, M~17, W~3, W~4, the Carina nebula, the Trifid nebula, NGC~3576, and NGC~1893, from which a sample of 31,784 MYStIX probable complex members \citep[MPCMs;][]{Broos2013-209} was obtained.
This catalog of 31,784 young stars includes both high- and low-mass stars and disk-bearing and disk-free stars \citep{Broos2013-209}. 
Although the \mystix\ samples are not complete, the young stars were identified in a uniform way for the different regions. This allowed a comparative analysis of the stellar populations in various regions. A comparison of the properties of a large cluster sample spanning a range of environments and ages enables inferences about processes such as gas removal, cluster expansion, subcluster mergers, dynamical boundedness and relaxation, and cluster dispersal, which all affect the spatial structure of clusters \citep{Kuhn2014-787,Kuhn2015-802,Kuhn2015-812}.

In 17 of the MYStIX MSFRs, \citet{Kuhn2014-787} identified 142 subclusters of young stars using finite-mixture models. Their physical parameters were summarized in \citet{Kuhn2015-812}. \citet{Kuhn2015-802} presented a census of the total intrinsic populations (estimated total numbers of OB and pre-main-sequence stars down to 0.1~$M_\odot$) and the surface density distributions for these 17 MSFRs using catalogs of young stars. 
The observed surface densities for X-ray selected MPCMs were calculated using an adaptive-smoothing method and were then corrected to the intrinsic populations by dividing the observed surface densities by the detection fraction maps. These stellar surface density maps can thus be directly compared with each other. 
However, the structures observed in the surface density maps represent the projected distributions of stars within star-forming regions, leading to possible overlaps of physically separate star groups on the map. Additionally, some of the small dense subclusters might not be visible in the surface density maps. More details about these aspects are given in \citet{Kuhn2015-802}.

\section{Method}

Observations of MSFRs at various stages of their star-forming lifespan unveil diverse stellar cluster structures. Notably, a striking feature of the spatial distributions of young stars is their organization into clusters and subclusters. In addition to the main clusters within these regions, smaller stellar groups can be observed outside the main clusters, or even as subclusters within the main clusters
\citep{Lada2003-41,Portegies2010-48,Krumholz2019-57}.
Stars form from hierarchically collapsing molecular clouds, which often results in clustered star formation that mirrors the intricate structure of the original molecular cloud. In a similar way to what \citet{Zhou2024-682-173,Zhou2024-682-128} reported for molecular clouds and clumps, we used the dendrogram algorithm to decompose the surface density distributions of stars into hierarchical structures. This revealed the multiscale structures of the star clusters.


\section{Results and discussion}

\subsection{Dendrogram structures}\label{Dendrogram}

\begin{figure}
\centering
\includegraphics[width=0.48\textwidth]{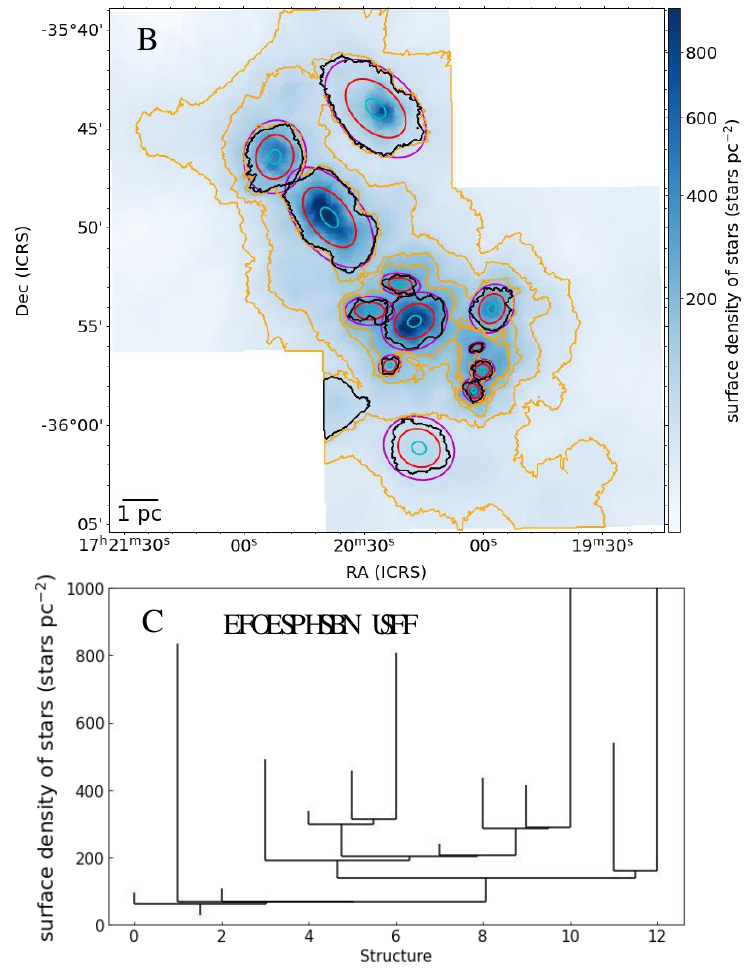}
\caption{Hierarchical structures of the clusters in NGC 6334. (a) The background is the surface density of stars. The black and orange contours show the masks of leaf and branch clusters, respectively. The magenta, red, and cyan cycles are the approximated ellipses with effective radii $R\rm_{eff}$, $R\rm_{eff,1}$, and $R\rm_{eff,2}$ defined in Sec.\ref{self}, respectively. (b) The dendrogram tree shows the hierarchical cluster structures.}
\label{example}
\end{figure}
\begin{figure}
\centering
\includegraphics[width=0.45\textwidth]{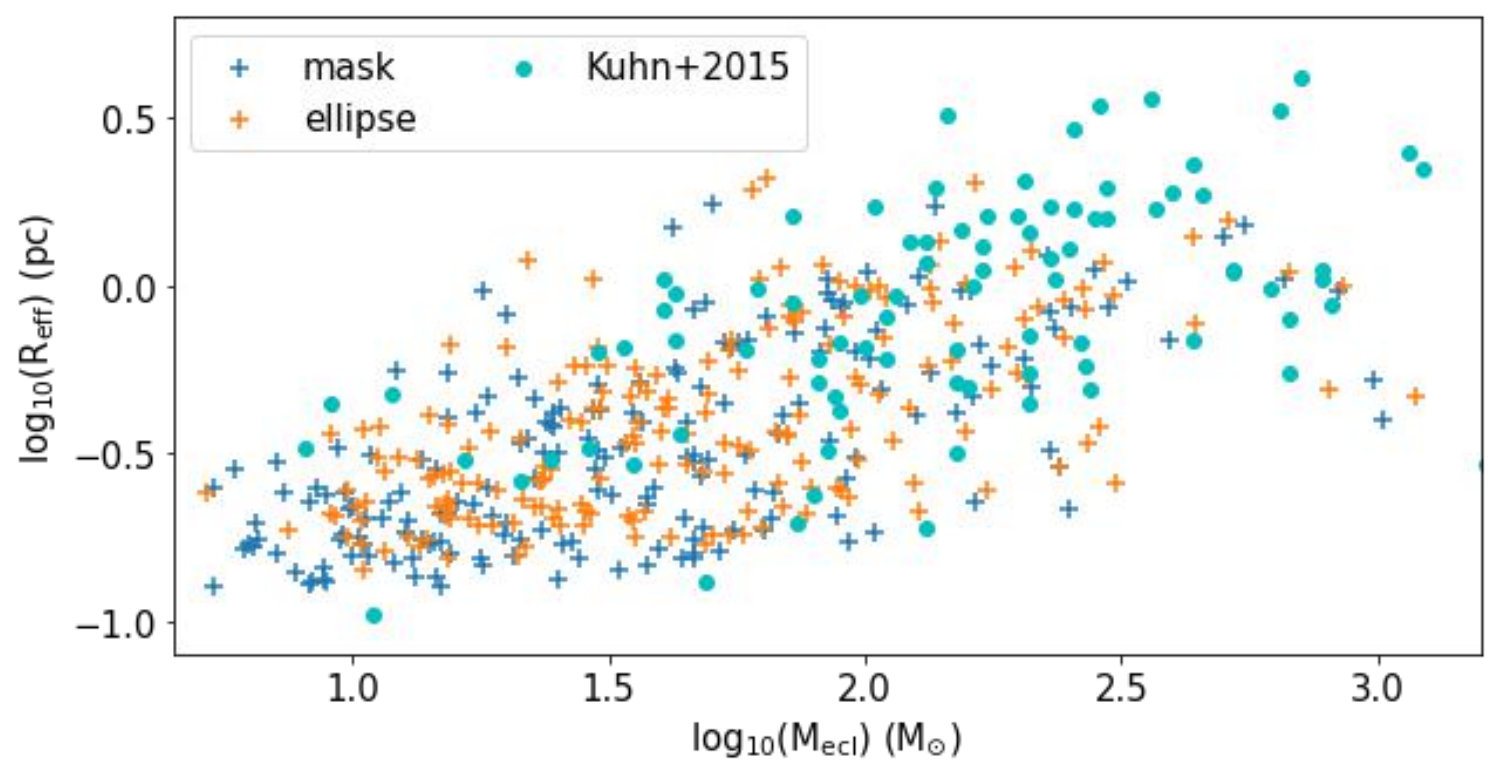}
\caption{Leaf clusters with the radii and masses calculated by the original mask and the elliptical mask (blue and orange pluses). The cyan dots show the subclusters in Table.1 of \citet{Kuhn2015-812}.}
\label{leaf}
\end{figure}

We identified local embedded star clusters using the dendrogram algorithm according to the surface density distributions of stars derived in \citet{Kuhn2015-802}. As described in \citet{Rosolowsky2008-679}, the dendrogram algorithm can decompose the density maps into hierarchical structures called branches and leaves (here, clusters and their subclusters). The {\it astrodendro} package \footnote{\url{https://dendrograms.readthedocs.io/en/stable/index.html}} includes three main input parameters for the dendrogram algorithm: {\it min\_value} for the minimum value to be considered in the dataset, {\it min\_delta} for a leaf that can be considered as an independent entity, and {\it min\_npix} for the minimum area of a structure.

In some observations, the cluster size was measured by the length scale over which the density profile exceeds twice the standard deviation of the surface density in the surrounding field \citep{Lada1991-371, Carpenter2000-120, Kumar2006-449}. According to the surface density map, we estimated the mean value of the low-density regions located at the edge as the background density, that is, $\Sigma_{\rm rms}$, then we set the parameters {\it min\_value}=2*$\Sigma_{\rm rms}$ and {\it min\_delta}=2*$\Sigma_{\rm rms}$. The identification is not sensitive to these two parameters. Taking 5*$\Sigma_{\rm rms}$ does not change the results significantly.
The number of the extracted structures in the two cases are very similar because the stellar surface density of local subclusters is much higher than the background stellar surface density ($\Sigma_{\rm rms}$).
The mass of the smallest stellar group observed in \citet{Kuhn2015-812} is $\approx 5 M_{\odot}$, according to the initial mass-radius relation of embedded clusters deduced by \citet{Marks2012-543}, that is, 
\begin{equation}
 \frac{r_{\rm h}}{{\rm pc}}=0.10_{-0.04}^{+0.07}\times\left(\frac{M_{\rm ecl}}{M_{\odot}}\right)^{0.13\pm0.04}\;,
 \label{rm}
\end{equation}
where $r_{\rm h}$ and $M_{\rm ecl}$ are the half-mass radius and the mass in stars of the embedded cluster, and
the corresponding minimum radius $r_{\rm min}$ of an embedded cluster is $\approx$0.12 pc. The minimum area can be calculated by $\pi r_{\rm min}^2$, which is used to fix the parameter {\it min\_npix}.
 Figure \ref{example} displays an example that shows the hierarchical structures of the clusters in NGC 6334, decomposed by the dendrogram algorithm.


\subsection{Cluster mass}

The {\it astrodendro} package can output the mask of each structure, which can be directly used to calculate the physical quantities of the identified structures. The area of each pixel is $A_{0}$ (unit: pixel$^{2}$), and the number of pixels included in a mask is $n$. The mass of a cluster is then

\begin{equation}
M = \sum_{i=1}^{n} \Sigma_{\rm i} A_{0}*0.5,
\label{mass}
\end{equation}where we assumed that the average mass of stars in a cluster is 0.5 M$_{\odot}$. $\Sigma_{\rm i}$ is the surface density of stars in a pixel (unit: n$_{s}$/pixel$^{2}$), with n$_{s}$ the number of stars per pixel.
The effective radius of the cluster is given by $R\rm_{eff}$ = $n*A_{0}/\pi$.

The algorithm can also approximate the morphology of each structure as an ellipse. In the dendrogram algorithm, the long and short axes of an ellipse, $a$ and $b$, are the {\it rms} sizes in radians (second moments) of the density distribution along the two spatial dimensions. However, as shown in Fig.\ref{example}, $a$ and $b$ give an ellipse that is much smaller than the size of the identified structure, but multiplying by a factor of 2.5 is appropriate to cover the structure \citep{Zhou2024-682-173,Zhou2024-682-128}. The effective physical radius of an ellipse is then $ R\rm_{eff}$ =$\sqrt{2.5a \times 2.5b}*d$, with $d$ being the distance to each star-forming region. The elliptical mask can also be used to calculate the mass using Eq.~\ref{mass}.

Fig.~\ref{leaf} shows that similar calculation results are obtained with the original or the elliptical mask. Moreover, the radii and masses of leaf clusters are systematically smaller and lower than those of the subclusters in Table.1 of \citet{Kuhn2015-812}. Thus, they systematically move down to the lower left side.
In Fig.~\ref{example}(a), the branch clusters present a complex morphology, 
and thus, only the original mask was used to do the calculations.

\subsection{Self-similar mass-size relations}\label{self}

\begin{figure}
\centering
\includegraphics[width=0.48\textwidth]{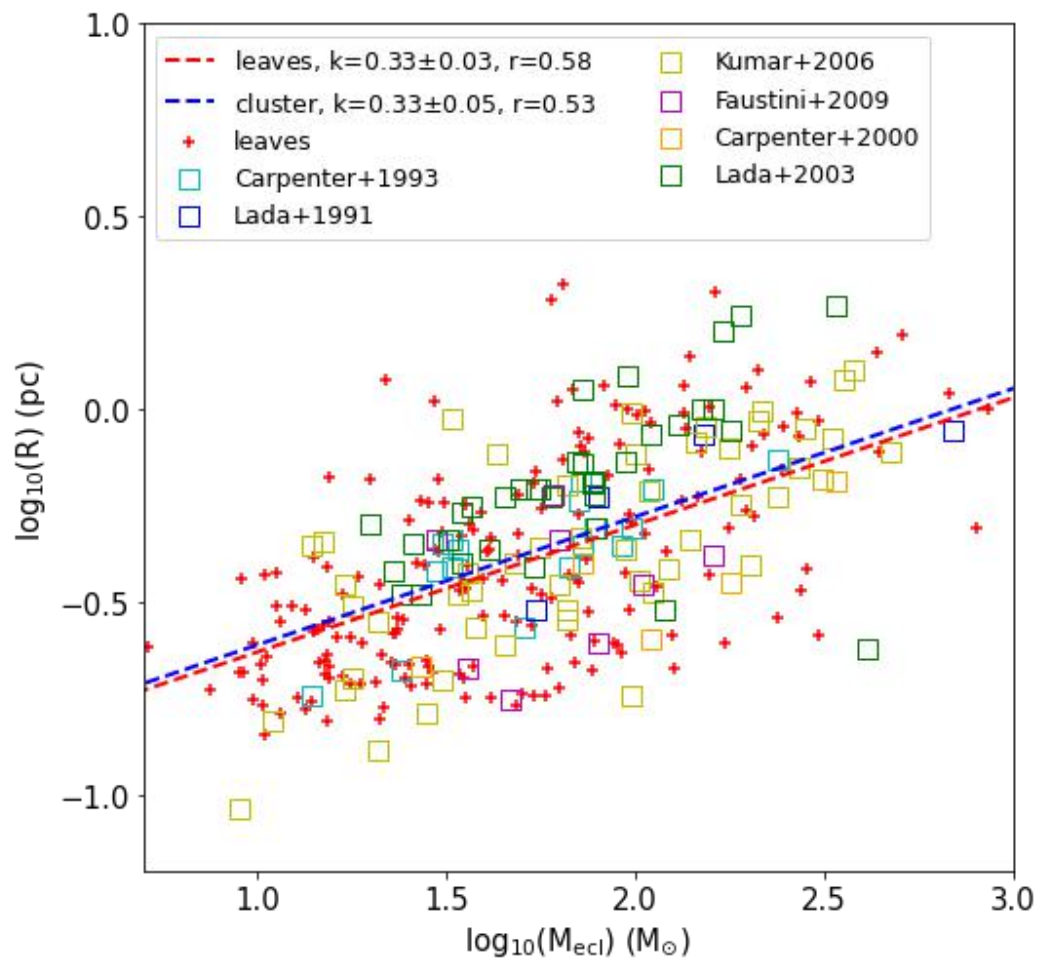}
\caption{Mass-size relations of leaf clusters (red plus) identified in this work fit by the dashed red line, and embedded clusters (colored squares) in the catalogs of \citet{Lada1991-371, Carpenter1993-407, Lada2003-41, Carpenter2000-120, Kumar2006-449, Faustini2009-503} fit by the dashed black line. The slope for all is $\approx$0.33.}
\label{Zhou}
\end{figure}

\begin{figure}
\centering
\includegraphics[width=0.45\textwidth]{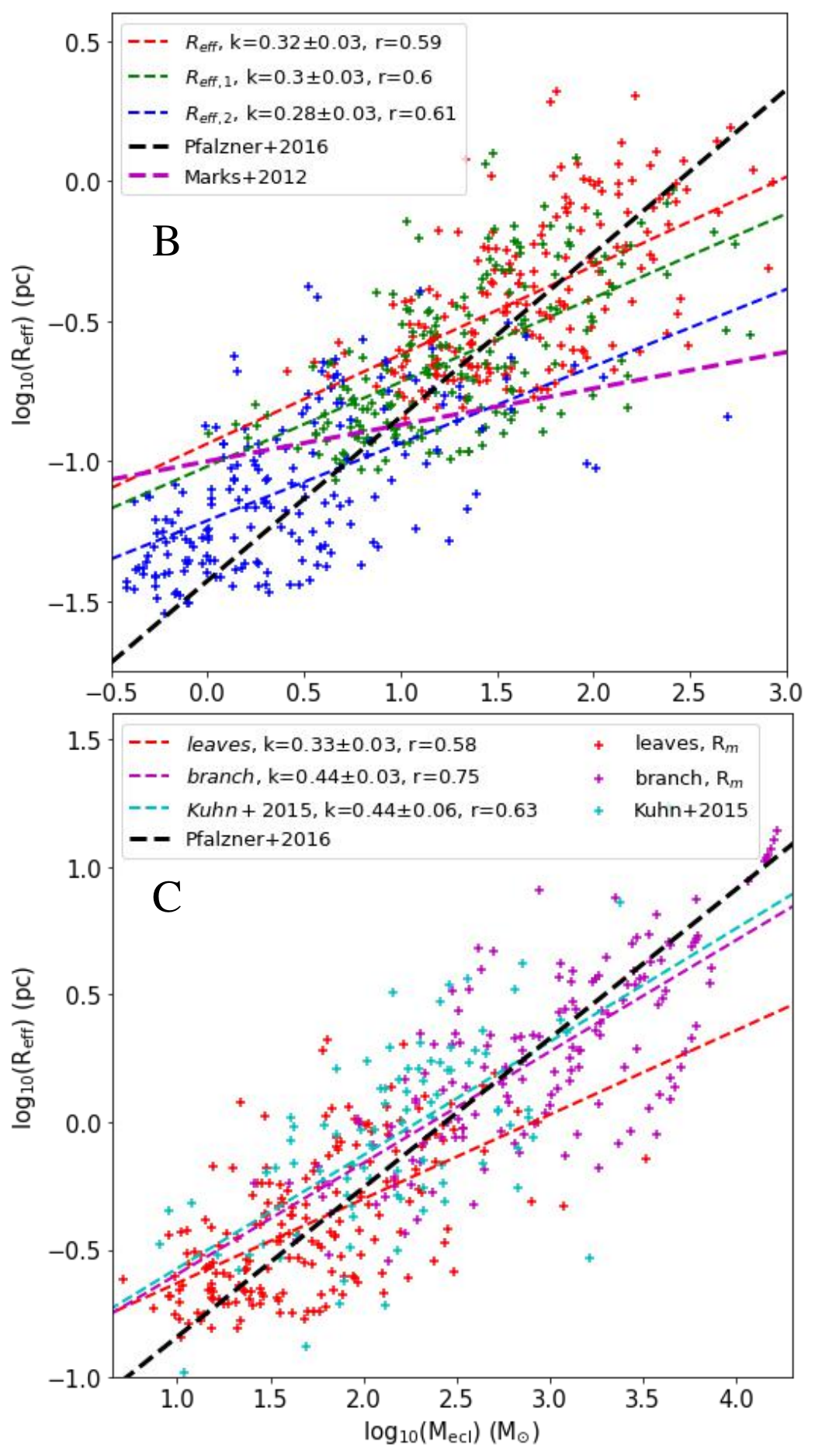}
\caption{ Mass-size relation of star clusters measured at different scales.
(a) Red, green, and blue pluses represent leaf clusters with effective radii $R\rm_{eff}$, $R\rm_{eff,1}$, and $R\rm_{eff,2}$ defined in Sec.\ref{self} and marked by colored ellipses in Fig.\ref{example}(a), respectively. (b) Magenta, cyan, and red pluses represent branch clusters, which are the subclusters in Table.1 of \citet{Kuhn2015-812}, and leaf clusters, respectively.
$k$ and $r$ are the slope of the mass-size relation and the Pearson correlation coefficient, respectively.}
\label{compare}
\end{figure}

Fig.~\ref{Zhou} shows that after the observed radii are unified to the half-mass radii \citep{Zhou2024sub},
the embedded clusters in the catalogs of \citet{Lada1991-371, Carpenter1993-407, Lada2003-41, Carpenter2000-120, Kumar2006-449, Faustini2009-503} are quite comparable with the leaf clusters identified in this work. Their mass-size relations have almost the same slope, $\approx$0.33.

As marked in Fig.~\ref{example}(a) by colored ellipses,
we calculated the physical parameters of leaf clusters by taking the effective radius of the ellipse as $ R_{\rm eff}$ =$\sqrt{2.5a \times 2.5b}*d$, $ R_{\rm eff, 1}$ =$\sqrt{1.5a \times 1.5b}*d$, and $R_{\rm eff, 2}$ =$\sqrt{0.5a \times 0.5b}*d$. As shown in Fig.~\ref{compare}(a), for these three cases, the data points mainly move down to the lower left side along the line of the mass-size relation with a constant slope of $\approx$0.3. 
Therefore, the slope of the mass-size relation is independent of the cluster radius and mass measurements, implying self-similar cluster structures. Interestingly, when the three different radii and mass measurements are mixed together, a steeper mass-size relation is obtained, similar to the mass-size relation found by \citet{Pfalzner2016-586}. 

Branch clusters are the complexes of leaf clusters. As shown in Fig~\ref{compare}(b), they follow a similar trend as those shown in Fig~\ref{compare}(a). When leaf and branch clusters were mixed atogether, we again obtained the steeper mass-size relation.
The data points in Table 1 of \citet{Kuhn2015-812} also follow this trend, although they were identified and measured with completely different methods. The radii and masses of the subclusters in \citet{Kuhn2015-812} are quite comparable with the middle-scale branch clusters in this work, and they have similar slopes for the mass-size relations in Fig.\ref{compare}(b).

In conclusion, different measurements for embedded clusters at different scales and ages give comparable mass-size relations. 
Here and in \citet{Kuhn2015-812}, a unified measurement of the samples results in a shallower slope for the mass-size relation. In contrast, when different measurements at different scales are mixed, the mass-size relation is always steeper. 
Although the mixing may not make sense, it is unavoidable when different catalogs from different observations are collected together. Thus, in \citet{Zhou2024sub}, we unified the observed radii from different observations to the half-mass radii using the templates obtained from direct N-body simulations. In this work, the unified measurement of the samples from the MYStIX project and the unified identification of the clusters effectively avoids this issue.

The exponent $k\approx$0.33 in Fig.~\ref{Zhou} and Fig.~\ref{compare}(a) indicates a uniform distribution of the mass through the scale if the structures were to be roundish. This is consistent with the suggestions made in \citet{Zhou2024sub}, namely that all subclusters initially have a Plummer profile. The Plummer radius serves as a measure of the core radius, which is the region where the density remains relatively constant. For a Plummer profile with the Plummer radius $r_{\rm pl}$, the mass within the radius $r$ is given by
\begin{equation}
M(r)=M_{\rm ecl}\frac{\left(\frac{r}{r_{\rm pl}}\right)^3}{\left[1+\left(\frac{r}{r_{\rm pl}}\right)^2\right]^\frac{3}{2}}.
\end{equation}
Thus, different radii in different observations can be unified to the half-mass radii. This operation does not change the slope of the mass-size relation, which only shifts the line of the mass-size relation, as shown in Fig.\ref{compare}(a).

\subsection{Isolated evolution}

\begin{table}
\centering
\caption{Median spacing between clusters, between clumps, and between clusters and clumps within each massive star-forming region.}
\label{tab1}
\begin{tabular}{cccc}
\hline
Regions &       cluster (pc)    &       clump (pc)      &       cluster\&clump (pc)    \\
Lagoon  &       0.84    &       1.04    &       0.82    \\
NGC6334 &       1.34    &       0.82    &       0.75    \\
NGC6357 &       1.25    &       0.78    &       0.76    \\
Eagle   &       0.76    &       1.41    &       0.72    \\
M17     &       0.66    &       1.02    &       0.66    \\
Carina  &       1.86    &       2.36    &       1.42    \\
Trifid  &       1.67    &       2.19    &       1.67    \\
Mean value      &       1.20    &       1.37    &       1.00    \\
\hline
\label{para}
\end{tabular}
\end{table}

\begin{figure}
\centering
\includegraphics[width=0.48\textwidth]{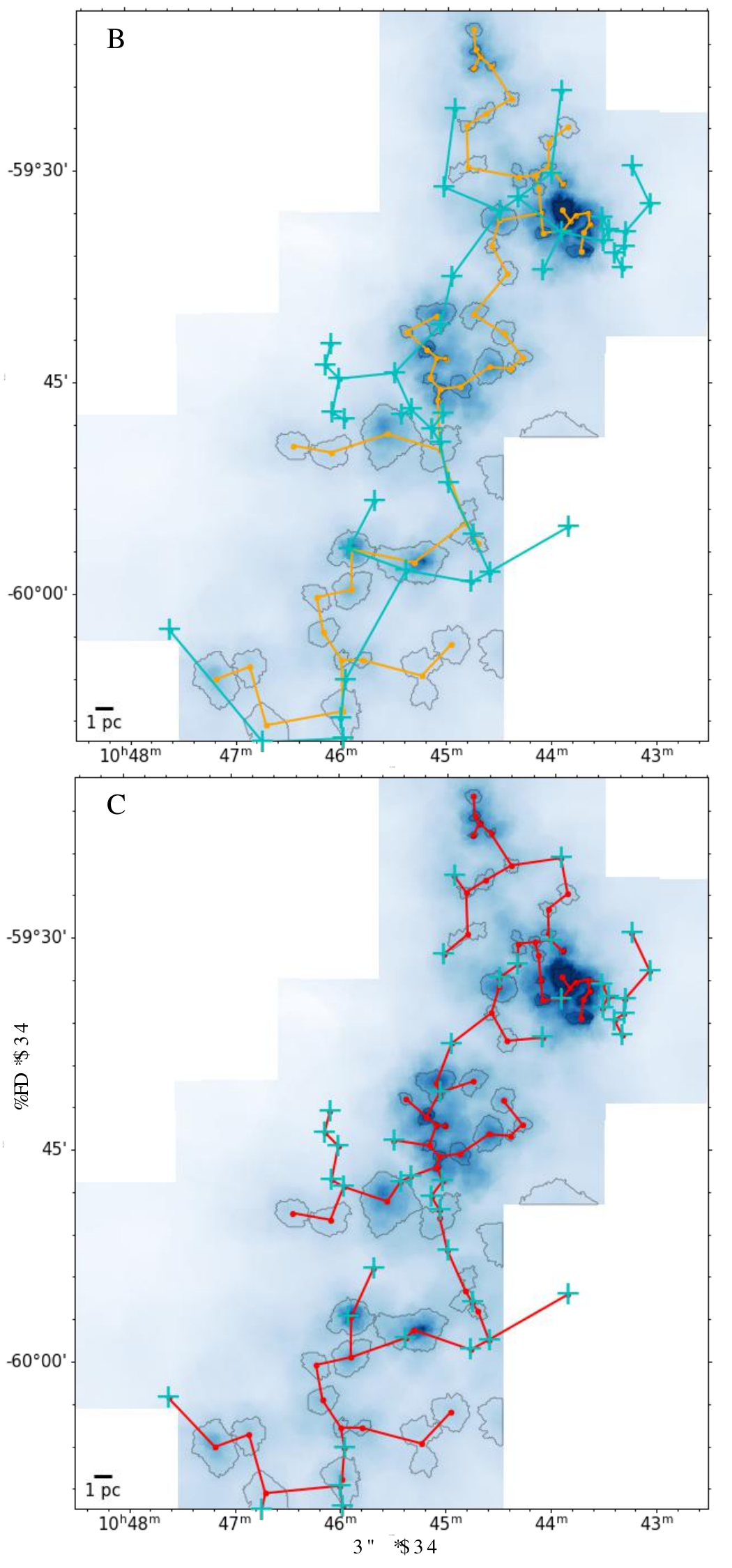}
\caption{
Carina nebula illustrates as an example the minimum spanning tree (MST) method.
The cyan plus marks the positions of ATLASGAL clumps. The black contours show the masks of leaf clusters. 
(a) Orange and cyan lines show the minimum spanning trees between clusters and between clumps, respectively. (b) MST between clusters and clumps.}
\label{mst}
\end{figure}

\begin{figure}
\centering
\includegraphics[width=0.48\textwidth]{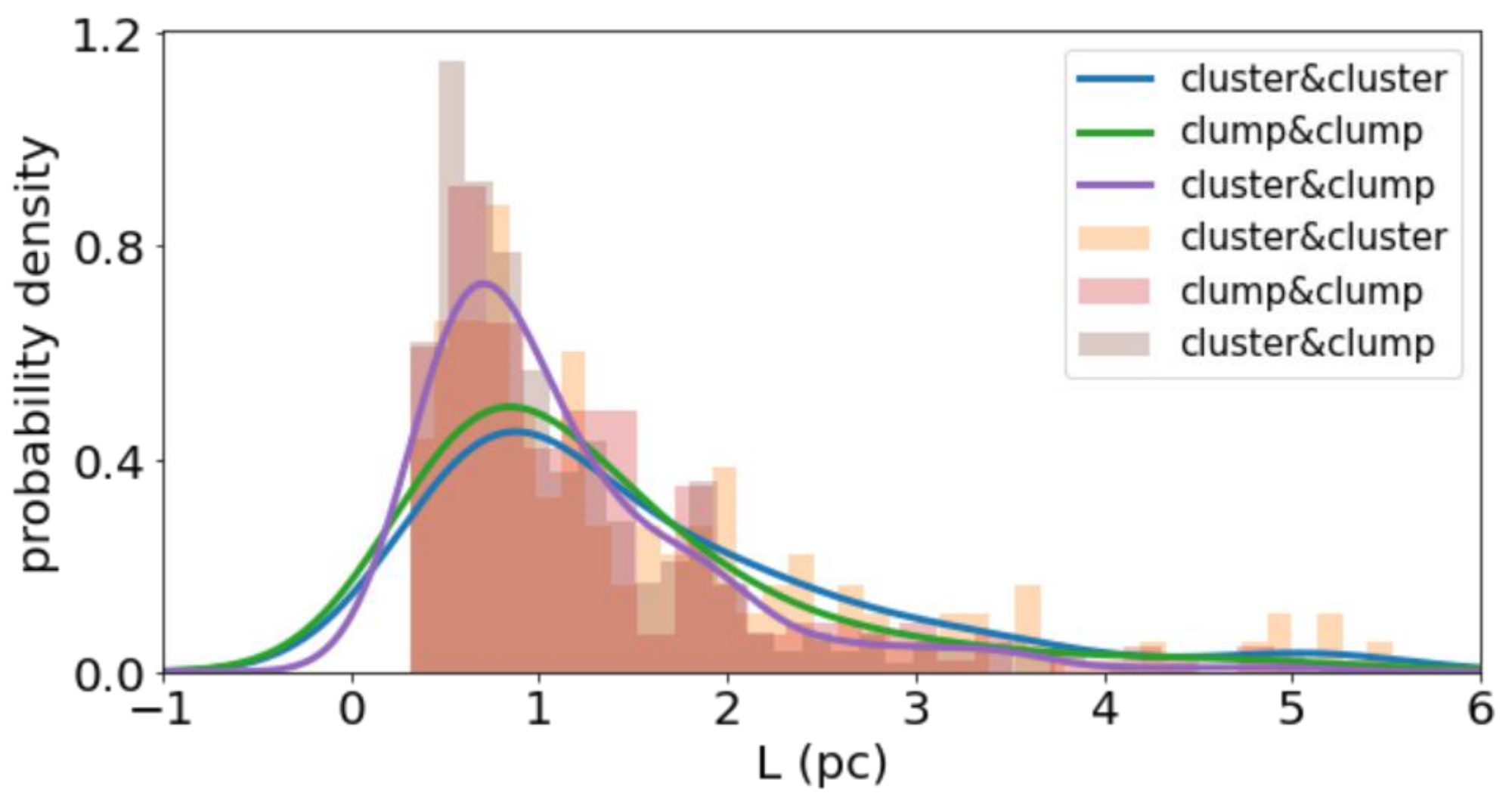}
\caption{Separation distributions of each two neighboring objects in three cases. The peak values for all are similar and close to 1 pc.}
\label{separation}
\end{figure}

\begin{figure}
\centering
\includegraphics[width=0.48\textwidth]{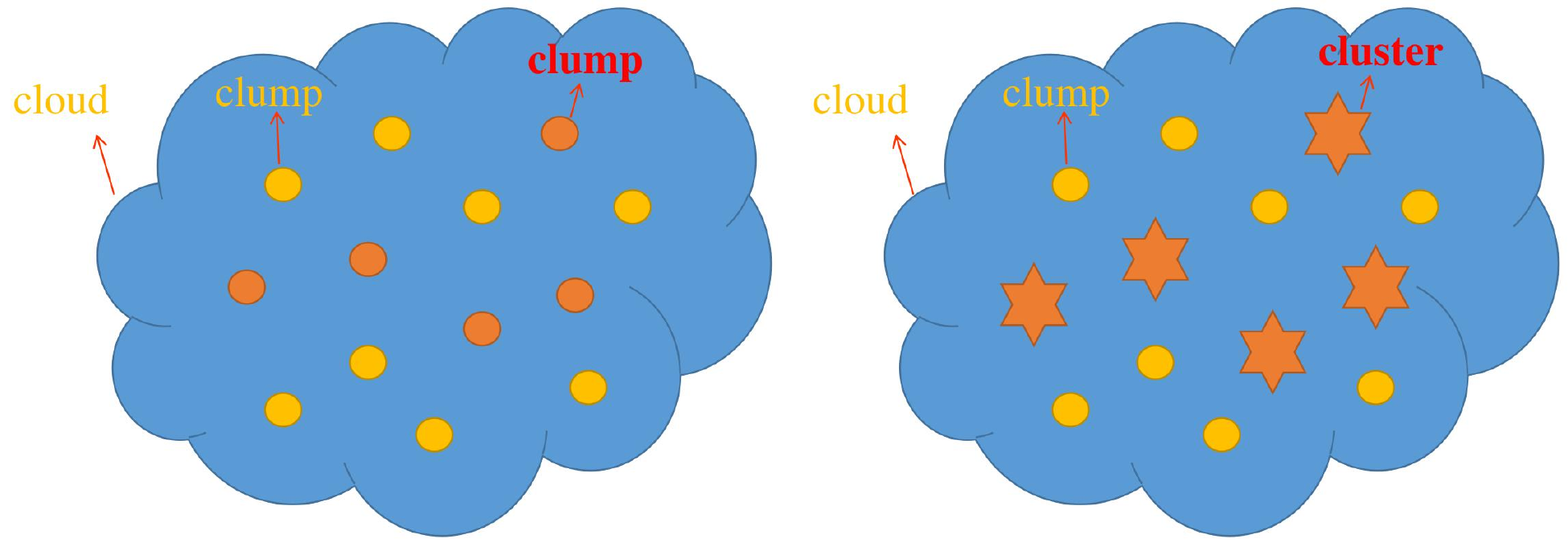}
\caption{Evolution from clump to embedded cluster fixed at the initial position of the clump.}
\label{carton}
\end{figure}

It remains controversial whether massive clusters are formed by the merger of clusters with lower masses. If there is a considerable merger between embedded clusters, we expect the separations between embedded clusters to be significantly smaller than their precursors, that is, clumps. In order to test this hypothesis, we used the clumps from the ATLASGAL survey, which provides the most well-characterized currently available sample of high-mass star-forming clumps \citep{Urquhart2022-510}. Seven of the 17 MSFRs are covered in the ATLASGAL survey, which are the Lagoon nebula, NGC~6334, NGC~6357, the Eagle nebula, M~17, the Carina nebula, and the Trifid nebula. We only considered the clumps contained in the fields of view of the MYStIX project. We also filtered out clumps whose kinematic distances differed significantly from the distance of the corresponding massive star-forming region. In Fig.~\ref{mst}, the minimum spanning tree (MST) method \citep{Cartwright2004-348,Dib2019-629A} was used to measure the separation between each two neighboring objects (between clusters, between clumps, and between clusters and clumps). 
Interestingly, in Tab.~\ref{para} and Fig.~\ref{separation}, the peak values of the separation distributions in three cases are comparable. This indicates the following:

1. The evolution from clump to embedded cluster is fixed at the initial position of the clump. There is no significant positional change between the embedded cluster and its precursor, as illustrated in Fig.~\ref{carton}. 

2. There is no significant merger between embedded clusters for the samples in the MYStIX project.

3. Feedback from embedded clusters does not redistribute the surrounding clumps, 
consistent with the conclusions in \citet{Zhou2024-682-173,Zhou2024-682-128}:
Feedback from embedded clusters does not significantly change the physical properties of the surrounding dense gas structures. 

In short, the evolution from a clump to an embedded cluster proceeds in isolation and locally, with no significant effect on the surrounding objects. 
Neither does the well-established $m_{\rm max}-M_{\rm ecl}$ relation \citep{Larson2003-287,Bonnell2004-349,Weidner2006-365,Weidner2013-434,Stephens2017-834,Yan2023-670}, a correlation between the embedded star cluster mass $M_{\rm ecl}$ and the mass of the most massive star $m_{\rm max}$, support the hypothesis that mergers during the formation of embedded clusters are common. Using Gaia DR2 data, \citet{Kuhn2019-870} investigated the kinematics of some subclusters identified in \citet{Kuhn2014-787}, They found no evidence that these groups are merging. We note that the imbalanced populations of star clusters and clumps in terms of numbers may impact the statistics of the spacing. More samples are necessary for further validation.

\subsection{Star formation efficiency}

\begin{figure}
\centering
\includegraphics[width=0.48\textwidth]{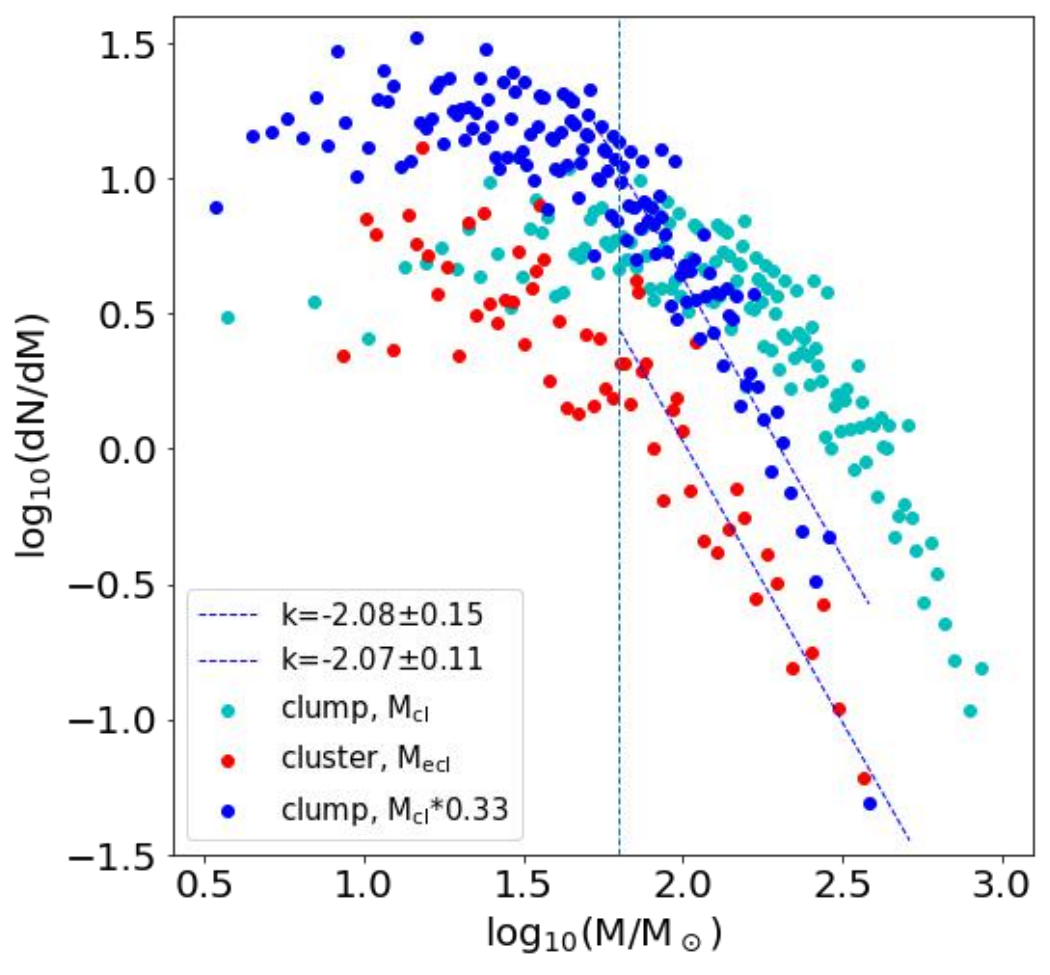}
\caption{Cluster and clump mass functions. The embedded clusters are taken from Fig.\ref{Zhou}. The vertical dashed line marks the start point of the linear fit. k is the slope of the fitting.}
\label{cmf}
\end{figure}

The massive star-forming regions in the MYStIX project analyzed in \citet{Kuhn2014-787} are located at distances < 3.6 kpc. We also adopted this distance criterion to select the ATLASGAL clumps in \citet{Urquhart2022-510}, which were used for a comparison with the identified embedded clusters. This distance limitation can also effectively eliminate a possible distance bias \citep{Urquhart2022-510}. 

Clumps are the precursors of embedded clusters.
\citet{Urquhart2022-510} classified 5007 ATLASGAL clumps  into four evolutionary stages, that is, quiescent, protostellar, young stellar objects (YSOs), and HII regions. 
The ATLASGAL clumps with HII regions (HII-clumps) represent the final stages of embedded cluster formation in the clumps.
When we consider the hub-filament structure of the clump in high-resolution ALMA observations \citep{Zhou2022-514,Zhou2024-686}, the hub is the true formation site of an embedded cluster within the clump. Thus,
the size of the clump should be significantly larger than the initial radius of the embedded cluster generated in the clump, which is also indicated by the initial mass--radius relation of embedded star clusters of \citet{Marks2012-543}.
Therefore, the mass-size relation of the clumps is not comparable with that of the embedded clusters. The star formation efficiency (SFE) cannot be inferred from the differences between the mass-size relations of the clumps and embedded clusters, as was done in \citet{Pfalzner2016-586}.
However, before strong feedback occurs, the total mass of a clump at different evolutionary stages remains almost constant. The SFE can be deduced from the mass functions of the clumps and embedded clusters.

A constant star formation efficiency (SFE) of $\approx$0.33 is revealed in Fig.~\ref{cmf}. This value has been widely used in previous simulations and has been proven to be effective \citep{Kroupa2001-321,
Banerjee2012-746,
Banerjee2013-764,
Banerjee2014-787,
Banerjee2015-447,
Oh2015-805,
Oh2016-590,
Banerjee2017-597,
Brinkmann2017-600,
Oh2018-481,
Wang2019-484,
Pavlik2019-626,
Wang2020-491,
Dinnbier2022-660}. 
This SFE is also consistent with the value obtained from
hydrodynamic calculations including self-regulation \citep{Machida2012-421,Bate2014-437} and also with observations of embedded systems in the solar neighborhood \citep{Lada2003-41,Megeath2016-151}.

\section{Conclusion}

We used the dendrogram algorithm to decompose the surface density distributions of stars from the MYStIX project into hierarchical structures. The algorithm presents multiscale structures of star clusters.
Our main conclusions are listed below.\\

1. Different measurements of embedded clusters at different scales and ages give comparable mass-size relations. The simple explanation is that all embedded clusters possess a Plummer profile. 

2. In a unified measurement of the same sample, the mass-size relation has a shallower slope. Instead, mixing different measurements at different scales together always yields a steeper mass-size relation. This issue cannot be avoided when different catalogs from different observations are collected together. The observed radii from different observations must be unified before an analysis.

3. We used the minimum spanning tree (MST) method to measure the separations between clusters and clumps in each massive star-forming region. The separations between cluster-cluster, clump-clump, and cluster-clump are comparable, which indicates that the evolution from a clump to an embedded cluster is fixed at the initial position of the clump. There is no significant positional change between the embedded cluster and its precursor. There is no significant merger between embedded clusters for the samples in the MYStIX project. Feedback from embedded clusters does not redistribute the surrounding clumps. However, more samples are necessary for a further validation.

4. The mass-size relation of the clumps is not comparable with that of the embedded clusters. However, by comparing the mass functions of the ATLASGAL clumps and the identified embedded clusters, we confirmed that a constant star formation efficiency of $\approx 0.33$ can be a typical value for the ATLASGAL clumps. 

\begin{acknowledgements}
We would like to thank the referee for the detailed comments and suggestions that improve and clarify this work.
\end{acknowledgements}

\bibliographystyle{aa} 
\bibliography{ref}

\begin{thebibliography}{54}
\expandafter\ifx\csname natexlab\endcsname\relax\def\natexlab#1{#1}\fi

\bibitem[{{Banerjee} \& {Kroupa}(2013)}]{Banerjee2013-764}
{Banerjee}, S. \& {Kroupa}, P. 2013, \apj, 764, 29

\bibitem[{{Banerjee} \& {Kroupa}(2014)}]{Banerjee2014-787}
{Banerjee}, S. \& {Kroupa}, P. 2014, \apj, 787, 158

\bibitem[{{Banerjee} \& {Kroupa}(2015)}]{Banerjee2015-447}
{Banerjee}, S. \& {Kroupa}, P. 2015, \mnras, 447, 728

\bibitem[{{Banerjee} \& {Kroupa}(2017)}]{Banerjee2017-597}
{Banerjee}, S. \& {Kroupa}, P. 2017, \aap, 597, A28

\bibitem[{{Banerjee} {et~al.}(2012){Banerjee}, {Kroupa}, \& {Oh}}]{Banerjee2012-746}
{Banerjee}, S., {Kroupa}, P., \& {Oh}, S. 2012, \apj, 746, 15

\bibitem[{{Bate} {et~al.}(2014){Bate}, {Tricco}, \& {Price}}]{Bate2014-437}
{Bate}, M.~R., {Tricco}, T.~S., \& {Price}, D.~J. 2014, \mnras, 437, 77

\bibitem[{{Bonnell} {et~al.}(2004){Bonnell}, {Vine}, \& {Bate}}]{Bonnell2004-349}
{Bonnell}, I.~A., {Vine}, S.~G., \& {Bate}, M.~R. 2004, \mnras, 349, 735

\bibitem[{{Brinkmann} {et~al.}(2017){Brinkmann}, {Banerjee}, {Motwani}, \& {Kroupa}}]{Brinkmann2017-600}
{Brinkmann}, N., {Banerjee}, S., {Motwani}, B., \& {Kroupa}, P. 2017, \aap, 600, A49

\bibitem[{{Broos} {et~al.}(2013){Broos}, {Getman}, {Povich}, {Feigelson}, {Townsley}, {Naylor}, {Kuhn}, {King}, \& {Busk}}]{Broos2013-209}
{Broos}, P.~S., {Getman}, K.~V., {Povich}, M.~S., {et~al.} 2013, \apjs, 209, 32

\bibitem[{{Carpenter}(2000)}]{Carpenter2000-120}
{Carpenter}, J.~M. 2000, \aj, 120, 3139

\bibitem[{{Carpenter} {et~al.}(1993){Carpenter}, {Snell}, {Schloerb}, \& {Skrutskie}}]{Carpenter1993-407}
{Carpenter}, J.~M., {Snell}, R.~L., {Schloerb}, F.~P., \& {Skrutskie}, M.~F. 1993, \apj, 407, 657

\bibitem[{{Cartwright} \& {Whitworth}(2004)}]{Cartwright2004-348}
{Cartwright}, A. \& {Whitworth}, A.~P. 2004, \mnras, 348, 589

\bibitem[{{Dib}(2023)}]{Dib2023-524}
{Dib}, S. 2023, \mnras, 524, 1625

\bibitem[{{Dib} \& {Henning}(2019)}]{Dib2019-629A}
{Dib}, S. \& {Henning}, T. 2019, \aap, 629, A135

\bibitem[{{Dinnbier} {et~al.}(2022){Dinnbier}, {Kroupa}, \& {Anderson}}]{Dinnbier2022-660}
{Dinnbier}, F., {Kroupa}, P., \& {Anderson}, R.~I. 2022, \aap, 660, A61

\bibitem[{{Faustini} {et~al.}(2009){Faustini}, {Molinari}, {Testi}, \& {Brand}}]{Faustini2009-503}
{Faustini}, F., {Molinari}, S., {Testi}, L., \& {Brand}, J. 2009, \aap, 503, 801

\bibitem[{{Feigelson} {et~al.}(2013){Feigelson}, {Townsley}, {Broos}, {Busk}, {Getman}, {King}, {Kuhn}, {Naylor}, {Povich}, {Baddeley}, {Bate}, {Indebetouw}, {Luhman}, {McCaughrean}, {Pittard}, {Pudritz}, {Sills}, {Song}, \& {Wadsley}}]{Feigelson2013-209}
{Feigelson}, E.~D., {Townsley}, L.~K., {Broos}, P.~S., {et~al.} 2013, \apjs, 209, 26

\bibitem[{{Kroupa}(1995{\natexlab{a}})}]{Kroupa1995a-277}
{Kroupa}, P. 1995{\natexlab{a}}, \mnras, 277, 1491

\bibitem[{{Kroupa}(1995{\natexlab{b}})}]{Kroupa1995b-277}
{Kroupa}, P. 1995{\natexlab{b}}, \mnras, 277, 1507

\bibitem[{{Kroupa}(2005)}]{Kroupa2005-576}
{Kroupa}, P. 2005, in ESA Special Publication, Vol. 576, The Three-Dimensional Universe with Gaia, ed. C.~{Turon}, K.~S. {O'Flaherty}, \& M.~A.~C. {Perryman}, 629

\bibitem[{{Kroupa} {et~al.}(2001){Kroupa}, {Aarseth}, \& {Hurley}}]{Kroupa2001-321}
{Kroupa}, P., {Aarseth}, S., \& {Hurley}, J. 2001, \mnras, 321, 699

\bibitem[{{Krumholz} {et~al.}(2019){Krumholz}, {McKee}, \& {Bland-Hawthorn}}]{Krumholz2019-57}
{Krumholz}, M.~R., {McKee}, C.~F., \& {Bland-Hawthorn}, J. 2019, \araa, 57, 227

\bibitem[{{Kuhn} {et~al.}(2014){Kuhn}, {Feigelson}, {Getman}, {Baddeley}, {Broos}, {Sills}, {Bate}, {Povich}, {Luhman}, {Busk}, {Naylor}, \& {King}}]{Kuhn2014-787}
{Kuhn}, M.~A., {Feigelson}, E.~D., {Getman}, K.~V., {et~al.} 2014, \apj, 787, 107

\bibitem[{{Kuhn} {et~al.}(2015{\natexlab{a}}){Kuhn}, {Feigelson}, {Getman}, {Sills}, {Bate}, \& {Borissova}}]{Kuhn2015-812}
{Kuhn}, M.~A., {Feigelson}, E.~D., {Getman}, K.~V., {et~al.} 2015{\natexlab{a}}, \apj, 812, 131

\bibitem[{{Kuhn} {et~al.}(2015{\natexlab{b}}){Kuhn}, {Getman}, \& {Feigelson}}]{Kuhn2015-802}
{Kuhn}, M.~A., {Getman}, K.~V., \& {Feigelson}, E.~D. 2015{\natexlab{b}}, \apj, 802, 60

\bibitem[{{Kuhn} {et~al.}(2019){Kuhn}, {Hillenbrand}, {Sills}, {Feigelson}, \& {Getman}}]{Kuhn2019-870}
{Kuhn}, M.~A., {Hillenbrand}, L.~A., {Sills}, A., {Feigelson}, E.~D., \& {Getman}, K.~V. 2019, \apj, 870, 32

\bibitem[{{Kumar} {et~al.}(2006){Kumar}, {Keto}, \& {Clerkin}}]{Kumar2006-449}
{Kumar}, M.~S.~N., {Keto}, E., \& {Clerkin}, E. 2006, \aap, 449, 1033

\bibitem[{{Lada} \& {Lada}(2003)}]{Lada2003-41}
{Lada}, C.~J. \& {Lada}, E.~A. 2003, \araa, 41, 57

\bibitem[{{Lada} {et~al.}(1991){Lada}, {Depoy}, {Evans}, \& {Gatley}}]{Lada1991-371}
{Lada}, E.~A., {Depoy}, D.~L., {Evans}, Neal~J., I., \& {Gatley}, I. 1991, \apj, 371, 171

\bibitem[{{Larson}(2003)}]{Larson2003-287}
{Larson}, R.~B. 2003, in Astronomical Society of the Pacific Conference Series, Vol. 287, Galactic Star Formation Across the Stellar Mass Spectrum, ed. J.~M. {De Buizer} \& N.~S. {van der Bliek}, 65--80

\bibitem[{{Machida} \& {Matsumoto}(2012)}]{Machida2012-421}
{Machida}, M.~N. \& {Matsumoto}, T. 2012, \mnras, 421, 588

\bibitem[{{Marks} \& {Kroupa}(2012)}]{Marks2012-543}
{Marks}, M. \& {Kroupa}, P. 2012, \aap, 543, A8

\bibitem[{{Megeath} {et~al.}(2016){Megeath}, {Gutermuth}, {Muzerolle}, {Kryukova}, {Hora}, {Allen}, {Flaherty}, {Hartmann}, {Myers}, {Pipher}, {Stauffer}, {Young}, \& {Fazio}}]{Megeath2016-151}
{Megeath}, S.~T., {Gutermuth}, R., {Muzerolle}, J., {et~al.} 2016, \aj, 151, 5

\bibitem[{{Motte} {et~al.}(2018){Motte}, {Bontemps}, \& {Louvet}}]{Motte2018}
{Motte}, F., {Bontemps}, S., \& {Louvet}, F. 2018, \araa, 56, 41

\bibitem[{{Oh} \& {Kroupa}(2016)}]{Oh2016-590}
{Oh}, S. \& {Kroupa}, P. 2016, \aap, 590, A107

\bibitem[{{Oh} \& {Kroupa}(2018)}]{Oh2018-481}
{Oh}, S. \& {Kroupa}, P. 2018, \mnras, 481, 153

\bibitem[{{Oh} {et~al.}(2015){Oh}, {Kroupa}, \& {Pflamm-Altenburg}}]{Oh2015-805}
{Oh}, S., {Kroupa}, P., \& {Pflamm-Altenburg}, J. 2015, \apj, 805, 92

\bibitem[{{Pavlik} {et~al.}(2019){Pavlik}, {Kroupa}, \& {{\v{S}}ubr}}]{Pavlik2019-626}
{Pavlik}, V., {Kroupa}, P., \& {{\v{S}}ubr}, L. 2019, \aap, 626, A79

\bibitem[{{Pfalzner} {et~al.}(2016){Pfalzner}, {Kirk}, {Sills}, {Urquhart}, {Kauffmann}, {Kuhn}, {Bhandare}, \& {Menten}}]{Pfalzner2016-586}
{Pfalzner}, S., {Kirk}, H., {Sills}, A., {et~al.} 2016, \aap, 586, A68

\bibitem[{{Portegies Zwart} {et~al.}(2010){Portegies Zwart}, {McMillan}, \& {Gieles}}]{Portegies2010-48}
{Portegies Zwart}, S.~F., {McMillan}, S. L.~W., \& {Gieles}, M. 2010, \araa, 48, 431

\bibitem[{{Rosolowsky} {et~al.}(2008){Rosolowsky}, {Pineda}, {Kauffmann}, \& {Goodman}}]{Rosolowsky2008-679}
{Rosolowsky}, E.~W., {Pineda}, J.~E., {Kauffmann}, J., \& {Goodman}, A.~A. 2008, \apj, 679, 1338

\bibitem[{{Stephens} {et~al.}(2017){Stephens}, {Gouliermis}, {Looney}, {Gruendl}, {Chu}, {Weisz}, {Seale}, {Chen}, {Wong}, {Hughes}, {Pineda}, {Ott}, \& {Muller}}]{Stephens2017-834}
{Stephens}, I.~W., {Gouliermis}, D., {Looney}, L.~W., {et~al.} 2017, \apj, 834, 94

\bibitem[{{Urquhart} {et~al.}(2022){Urquhart}, {Wells}, {Pillai}, {Leurini}, {Giannetti}, {Moore}, {Thompson}, {Figura}, {Colombo}, {Yang}, {K{\"o}nig}, {Wyrowski}, {Menten}, {Rigby}, {Eden}, \& {Ragan}}]{Urquhart2022-510}
{Urquhart}, J.~S., {Wells}, M.~R.~A., {Pillai}, T., {et~al.} 2022, \mnras, 510, 3389

\bibitem[{{Wang} {et~al.}(2019){Wang}, {Kroupa}, \& {Jerabkova}}]{Wang2019-484}
{Wang}, L., {Kroupa}, P., \& {Jerabkova}, T. 2019, \mnras, 484, 1843

\bibitem[{{Wang} {et~al.}(2020){Wang}, {Kroupa}, {Takahashi}, \& {Jerabkova}}]{Wang2020-491}
{Wang}, L., {Kroupa}, P., {Takahashi}, K., \& {Jerabkova}, T. 2020, \mnras, 491, 440

\bibitem[{{Weidner} \& {Kroupa}(2006)}]{Weidner2006-365}
{Weidner}, C. \& {Kroupa}, P. 2006, \mnras, 365, 1333

\bibitem[{{Weidner} {et~al.}(2013){Weidner}, {Kroupa}, \& {Pflamm-Altenburg}}]{Weidner2013-434}
{Weidner}, C., {Kroupa}, P., \& {Pflamm-Altenburg}, J. 2013, \mnras, 434, 84

\bibitem[{{Yan} {et~al.}(2023){Yan}, {Jerabkova}, \& {Kroupa}}]{Yan2023-670}
{Yan}, Z., {Jerabkova}, T., \& {Kroupa}, P. 2023, \aap, 670, A151

\bibitem[{{Zhou} {et~al.}(2024{\natexlab{a}}){Zhou}, {Dib}, {Juvela}, {Sanhueza}, {Wyrowski}, {Liu}, \& {Menten}}]{Zhou2024-686}
{Zhou}, J.~W., {Dib}, S., {Juvela}, M., {et~al.} 2024{\natexlab{a}}, \aap, 686, A146

\bibitem[{{Zhou} {et~al.}(2024{\natexlab{b}}){Zhou}, {Dib}, {Wyrowski}, {Liu}, {Li}, {Sanhueza}, {Juvela}, {Xu}, {Liu}, {Baug}, {Peng}, {Menten}, \& {Bronfman}}]{Zhou2024-682-173}
{Zhou}, J.~W., {Dib}, S., {Wyrowski}, F., {et~al.} 2024{\natexlab{b}}, \aap, 682, A173

\bibitem[{{Zhou} {et~al.}(2024{\natexlab{c}}){Zhou}, {Kroupa}, \& {Wu}}]{Zhou2024sub}
{Zhou}, J.~W., {Kroupa}, P., \& {Wu}, W.~J. 2024{\natexlab{c}}, The evolution of the mass–radius relation of expanding very young star clusters, submitted to MNRAS

\bibitem[{{Zhou} {et~al.}(2022){Zhou}, {Liu}, {Evans}, {Garay}, {Goldsmith}, {G{\'o}mez}, {V{\'a}zquez-Semadeni}, {Liu}, {Stutz}, {Wang}, {Juvela}, {He}, {Li}, {Bronfman}, {Liu}, {Xu}, {Tej}, {Dewangan}, {Li}, {Zhang}, {Zhang}, {Ren}, {Tatematsu}, {Shing Li}, {Won Lee}, {Baug}, {Qin}, {Wu}, {Peng}, {Zhang}, {Liu}, {Luo}, {Ge}, {Saha}, {Chakali}, {Zhang}, {Kim}, {Ristorcelli}, {Shen}, \& {Li}}]{Zhou2022-514}
{Zhou}, J.-W., {Liu}, T., {Evans}, N.~J., {et~al.} 2022, \mnras, 514, 6038

\bibitem[{{Zhou} {et~al.}(2024{\natexlab{d}}){Zhou}, {Wyrowski}, {Neupane}, {Barlach Christensen}, {Menten}, {Li}, \& {Liu}}]{Zhou2024-682-128}
{Zhou}, J.~W., {Wyrowski}, F., {Neupane}, S., {et~al.} 2024{\natexlab{d}}, \aap, 682, A128

\bibitem[{{Zinnecker} \& {Yorke}(2007)}]{Zinnecker2007}
{Zinnecker}, H. \& {Yorke}, H.~W. 2007, \araa, 45, 481

\end{thebibliography}


\appendix

\end{document}